\documentclass[aps,prb,twocolumn,groupedaddress,showpacs,amsmath,amssymb,amsfonts]{revtex4}

\usepackage{latexsym,amssymb}
\usepackage{graphicx,color,pstricks}
\usepackage{epsfig,epsf,bm}

\definecolor{gray}{rgb}{0.7,0.7,0.7}

\renewcommand{\v}[1]{\ensuremath{\mathbf{#1}}} 

\begin{document}

\title{Electrically tunable two-channel Kondo fixed points in helical liquids}

\author{Yu-Li Lee}
\email{yllee@cc.ncue.edu.tw} \affiliation{Department of Physics,
National Changhua University of Education, Changhua, Taiwan, Republic of China}

\author{Yu-Wen Lee}
\email{ywlee@thu.edu.tw} \affiliation{Department of Physics,
Tunghai University, Taichung, Taiwan, Republic of China}

\date{\today}

\begin{abstract}
We study a quantum dot coupled to two edge states of a quantum spin Hall insulator through electron tunnelings
in the presence of a Rashba spin-orbital interaction induced by an external electric field. We show that if the
electron interactions on the edge states are repulsive, there are two possible phases, depending on the Luttinger
liquid parameter $K$. For $1/2<K<1$, the low-temperature physics is controlled by a previously identified
two-channel Kondo fixed point. For the edge states with even stronger repulsive interactions, i.e. $1/4<K<1/2$,
the system reaches another phase at low temperatures, described by a new two-channel Kondo fixed point. This
phase is separated from the original one by a continuous phase transition upon varying the value of $K$ through
the external electric field. The corresponding critical point is described by a free Dirac fermion
backscattered by a local potential. We investigate the low-temperature properties associated with this new fixed
point and also discuss the scaling behaviors of the system at the critical point.
\end{abstract}


\maketitle

\section{introduction}

The experimental discovery\cite{Konig} of the quantum spin Hall (QSH) insulator in HgTe
quantum wells following its theoretical predictions\cite{BHZ} paved a new road in the study of topological phases.
The QSH state\cite{Qi_RMP} is a member of the topologically nontrivial states of matter with the symmetry protected
topological order.\cite{GuWen} These states of matter have a finite bulk gap and, in the mean time, support gapless
edge or surface excitations. For the QSH insulator, the edge states are helical in the sense that, on each edge,
there is a counter-propagating Kramers' pair of states with opposite spin polarizations. The stability of the helical
edge states against potential scattering is protected by the time-reversal (TR) symmetry.\cite{KM, BZ, KM2, WBZ}
Contrary to the case of simple potential scattering, the edge electrons can backscatter from the magnetic impurities
through spin exchange, and thus the TR symmetry can no longer protect the helical states from mixing. The one-channel
Kondo ($1$CK) effect of helical edge states was studied in Refs. \onlinecite{SI, MLOQWZ, TFM, ESSJ}.

In a recent work, a quantum dot (QD) coupled to two helical edge states was studied.\cite{LSLN} It is well known that
such a system realizes the usual $1$CK effect for non-interacting electrons on the edges.\cite{GR,NL} In Ref.
\onlinecite{LSLN}, it is shown that for weakly repulsive interacting electrons on the edges, with the Luttinger
liquid (LL) parameter $K<1$, the system is driven to a two-channel Kondo ($2$CK) fixed point. This result is
non-trivial since in the context of a QD coupled to two LL leads, a much stronger Coulomb repulsion ($K<1/2$) is
warranted in order to realize the $2$CK physics.

A more complete description of the $2$CK physics in a QD coupled to two helical edge states must take into account
the Rashba spin-orbital interaction because this interaction, which can be tuned by an external gate voltage, is a
built-in feature of a quantum well.\cite{Wi} Moreover, the HgTe quantum wells exhibit some of the largest known
Rashba couplings of any semiconductor heterostructures.\cite{Bu} In fact, it was found very recently\cite{ESSJ} that
the presence of a Rashba coupling has profound effects on both the Kondo temperature and the transport properties of
the helical liquids in the presence of a single magnetic impurity.

\begin{figure}
\begin{center}
 \includegraphics[width=1.0\columnwidth]{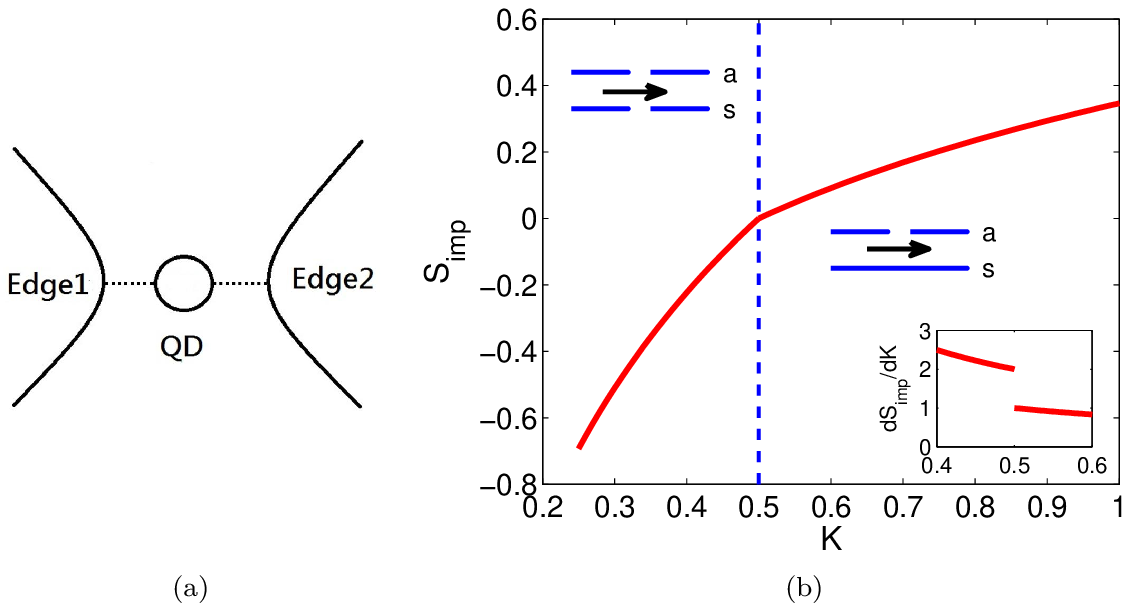}
 \caption{(Color online) %
 (a) A QD coupled to two helical edge states of QSH insulators in asymmetric HgTe/CdTe quantum wells through electron
 tunnelings. (b) The zero-temperature impurity entropy $S_{imp}$ versus the LL parameter $K$ in the range $1/4<K<1$. A
 singularity occurs at $K=1/2$, as shown by the inset which exhibits $dS_{imp}/dK$ versus $K$ around $K=1/2$. Moreover, we show
 the schematic pictures for the $2$CK ($1/2<K<1$) and $2$CK$^{\prime}$ ($1/4<K<1/2$) phases. The ground state of the system
 consists of two spinless wires---the symmetric sector ($s$) and the antisymmetric sector ($a$), and a partially screened spin (denoted by the dashed arrow). In the $2$CK phase, the antisymmetric sector is cut into two separated pieces while the symmetric sector is a spinless LL. In the $2$CK$^{\prime}$ phase, both sectors are cut into two separated pieces.}
 \label{2cksetup}
\end{center}
\end{figure}

In this work, we consider a QD coupled to two helical edge states in asymmetric HgTe/CdTe quantum wells, as shown
in Fig. \ref{2cksetup}(a). The Rashba coupling will be present in this system and we would like to investigate the effect of it on the $2$CK physics
studied in Ref. \onlinecite{LSLN}. We find that if the electron interactions on the edge states are repulsive,
there are two possible phases, depending on the value of $K$. For $1/2<K<1$, the low-temperature physics is controlled by the
$2$CK fixed point identified in Ref. \onlinecite{LSLN}. In this region, we find that the impurity entropy at zero temperature
$S_{imp}$ and the temperature dependence of the tunneling conductance between the two edges are not affected by the presence
of the Rashba coupling. On the other hand, the Rashba coupling reveals its presence through the temperature dependence of the
impurity specific heat $C_{imp}$ and the dynamical structure factor $S_i(\omega)$ of the spin in the dot at low temperatures,
i.e., $C_{imp}\propto T^{K+1/K-2}$ for $1/\sqrt{3}<K<1$, $C_{imp}\propto T^{4K-2}$ for $1/2<K<1/\sqrt{3}$, and $S_i(0)\propto
T^{K-1}$ with $i=y,z$. More interestingly, we show that the system reaches another phase at low temperatures, described by a
new $2$CK fixed point when $1/4<K<1/2$. In this region, $S_{imp}=\ln{(2K)}$, $C_{imp}\propto T^{1/K-2}$ or $T$, and $S_i(0)\propto
T^{2K-1}$. The two phases are separated by a quantum critical point (QCP), which is described by a free Dirac fermion
backscattered by a local potential. In the following, we outline the derivation of these results.

\section{Model}
We consider two helical edge states of a QSH insulator brought close to each other at a tunneling
junction. A QD is then placed at the middle of the junction and coupled to the two edge states through electron tunnelings.
When the number of electrons in the dot is odd and the repulsive interaction between the electrons in the dot is much
larger than the tunneling amplitude, the system can be described by the Kondo Hamiltonian $H=H_0+H_i+H_K$,\cite{LSLN, Hew}
where
\begin{equation}
 H_0=\! \sum_{m=1,2} \! \int \! \! dx \! \left(v_F\Psi_m^{\dagger}\sigma_3i\partial_x\Psi_m
 +\alpha\Psi_m^{\dagger}\sigma_2i\partial_x\Psi_m\right) , \label{ekh0}
\end{equation}
and
\begin{eqnarray}
 H_K \! \! \! &=& \! \! \! \sum_{m=1,2} \! \! \Psi^{\dagger}_m(0) \! \left(\sum_{i=x,y} \! \frac{J_{1\perp}}{2}S_i
 \sigma_i \! + \! \frac{J_{1z}}{2}S_z\sigma_z \! \right) \! \! \Psi_m(0) \nonumber \\
 \! \! \! & & \! \! + \! \sum_{m\neq n=1,2} \! \! \Psi^{\dagger}_m(0) \! \left(\sum_{i=x,y} \! \frac{J_{2\perp}}{2}S_i
 \sigma_i \! + \! \frac{J_{2z}}{2}S_z\sigma_z \! \right) \! \! \Psi_n(0) , \nonumber \\
 \! \! \! & & \! \! \label{ekh01}
\end{eqnarray}
In the above, $m=1,2$ label the two edges, $\Psi_m=[\psi_{m+},\psi_{m-}]^t$, $\psi_{m+}=\psi_{mL\uparrow}$,
$\psi_{m-}=\psi_{mR\downarrow}$, $\sigma=\uparrow,\downarrow=+,-$, and $\bm{S}$ is the spin operator for the spin-$1/2$
impurity. Here, $H_0$ and $H_i$ denote the kinetic energy (including the Rashba coupling with the strength $\alpha$)
and the electron Coulomb interaction of the helical edge states, respectively. $H_K$ describes the Kondo interaction
between electrons in the edges and a spin-$1/2$ magnetic impurity at $x=0$. The magnetic anisotropy, i.e.
$J_{l\perp}\neq J_{lz}$ with $l=1,2$, is induced by spin-orbital coupling.\cite{ZPP}

The Rashba term in $H_0$ can be absorbed into the kinetic term by a spinor rotation
$\tilde{\Psi}_m=e^{-i\frac{\theta}{2}\sigma_x}\Psi_m$,\cite{ESSJ, VO} where $\theta=\tan^{-1}{(\alpha/v_F)}$. By
rotating also the impurity spin $\tilde{\bm{S}}=e^{i\theta S_x}\bm{S}e^{-i\theta S_x}$, $H_0$ becomes
$
 \tilde{v}_F \! \sum_{m=1,2} \! \int \! \! dx~\tilde{\Psi}_m^{\dagger}\sigma_3i\partial_x
 \tilde{\Psi}_m \ ,
$
where $\tilde{v}_F=\sqrt{v_F^2+\alpha^2}$, and
\begin{eqnarray}
 H_K &=& \! \sum_{m=1,2}\sum_{i=x,y,z}\frac{\tilde{J}_{1i}}{2}\tilde{S}_i\tilde{\Psi}^{\dagger}_m
 \sigma_i\tilde{\Psi}_m(0) \nonumber \\
 & & +\! \sum_{m\neq n=1,2}\sum_{i=x,y,z}\frac{\tilde{J}_{2i}}{2}\tilde{S}_i\tilde{\Psi}^{\dagger}_m
 \sigma_i\tilde{\Psi}_n(0) \nonumber \\
 & & +\! \sum_{m=1,2}\frac{J_{1E}}{2}\tilde{\Psi}^{\dagger}_m(\tilde{S}_y\sigma_z+\tilde{S}_z
 \sigma_y)\tilde{\Psi}_m(0) \nonumber \\
 & & +\! \sum_{m\neq n=1,2}\frac{J_{2E}}{2}\tilde{\Psi}^{\dagger}_m(\tilde{S}_y\sigma_z
 +\tilde{S}_z\sigma_y)\tilde{\Psi}_n(0) \ , ~~~\label{hlkh13}
\end{eqnarray}
where $\tilde{J}_{lx}=J_{l\perp}$, $\tilde{J}_{ly}=J_{l\perp}\cos^2{\theta}+J_{lz}\sin^2{\theta}$, $\tilde{J}_{lz}=J_{lz}\cos^2{\theta}+J_{l\perp}\sin^2{\theta}$, and $J_{lE}=(J_{l\perp}-J_{lz})\sin{\theta}\cos{\theta}$
with $l=1,2$. Since the Rashba coupling respects the TR symmetry, we expect that the helical liquid on each edge is still described
by the LL. Hence, the allowed electron-electron interaction is of the form
\begin{equation}
 H_i=\! \sum_{m=1,2} \! \int \! \! dx\left(g_1 \! \sum_{\sigma} \! J_{m\sigma}J_{m\sigma}+g_2J_{m+}J_{m-}\right) , \label{hlkh12}
\end{equation}
where $J_{m\sigma}=\tilde{\psi}^{\dagger}_{m\sigma}\tilde{\psi}_{m\sigma}$.

To proceed, we bosonize the Kondo Hamiltonian $H=H_0+H_i+H_K$ according to the formula:
$\tilde{\psi}_{m\pm}=\frac{1}{\sqrt{2\pi a_0}}e^{\mp i\sqrt{4\pi}\phi_{m\pm}}$,\cite{GNT} where $a_0$ is the short-distance
cutoff. By defining the bosonic fields $\Phi_m=\phi_{m+}+\phi_{m-}$, $\Theta_m=\phi_{m+}-\phi_{m-}$, $\Phi_{s/a}=(\Phi_1\pm\Phi_2)/\sqrt{2}$,
and $\Theta_{s/a}=(\Theta_1\pm\Theta_2)/\sqrt{2}$, the Hamiltonian for the edge states can be written as
$H_0+H_i=\! \sum_{\alpha=s,a}\frac{v}{2} \! \int \! \! dx \! \left[K(\partial_x\Theta_{\alpha})^2+\frac{1}{K}(\partial_x\Phi_{\alpha})^2\right]$,
where the LL parameter $K$ and the speed of the collective excitation $v$ depend on both the Coulomb interaction and the
Rashba coupling strength. On the other hand, $H_K$ becomes
\begin{eqnarray}
 H_K &=& \frac{\tilde{J}_{1x}\tilde{S}_x}{\pi a_0}\cos{\! \left[\sqrt{2\pi}\Phi_s(0)\right]} \!
 \cos{\! \left[\sqrt{2\pi}\Phi_a(0)\right]} \nonumber \\
 & & +\frac{\tilde{J}_{1y}\tilde{S}_y+J_{1E}\tilde{S}_z}{\pi a_0}
 \sin{\! \left[\sqrt{2\pi}\Phi_s(0)\right]} \! \cos{\! \left[\sqrt{2\pi}\Phi_a(0)\right]}
 \nonumber \\
 & & +\frac{\tilde{J}_{2x}\tilde{S}_x}{\pi a_0}\cos{\! \left[\sqrt{2\pi}\Phi_s(0)\right]} \!
 \cos{\! \left[\sqrt{2\pi}\Theta_a(0)\right]} \nonumber \\
 & & +\frac{\tilde{J}_{2y}\tilde{S}_y+J_{2E}\tilde{S}_z}{\pi a_0}
 \sin{\! \left[\sqrt{2\pi}\Phi_s(0)\right]} \! \cos{\! \left[\sqrt{2\pi}\Theta_a(0)\right]}
 \nonumber \\
 & & -\frac{\tilde{J}_{2z}\tilde{S}_z+J_{2E}\tilde{S}_y}{\pi a_0}
 \sin{\! \left[\sqrt{2\pi}\Phi_a(0)\right]} \! \sin{\! \left[\sqrt{2\pi}\Theta_a(0)\right]}
 \nonumber \\
 & & +\frac{1}{\sqrt{2\pi}}(\tilde{J}_{1z}\tilde{S}_z+J_{1E}\tilde{S}_y)\partial_x\Theta_s(0) \ .
 \label{hlkh22}
\end{eqnarray}
In general, there are extra local backscattering terms caused by the QD in the Hamiltonian $H$. However, these terms are
irrelevant as long as $K>1/4$.\cite{WBZ} In the following, we shall focus on the regime with $1/4<K<1$, so that these local
backscattering terms can be neglected for low-energy physics.

\section{Scaling analysis}
Near the Gaussian fixed point ($\tilde{J}_{li}=0=J_{lE}$ with $l=1,2$ and $i=x,y,z$), the scaling dimensions of the various
terms in Eq.~(\ref{hlkh22}) are $\Delta(\tilde{J}_{1x})=K=\Delta(\tilde{J}_{1y})=\Delta(J_{1E})$, $\Delta (\tilde{J})_{1z}=1$,
and $\Delta(\tilde{J}_{2x})=K/2+1/(2K)=\Delta(\tilde{J}_{2y})=\Delta(\tilde{J}_{2z})=\Delta(J_{2E})$. Thus, for $K<1$, the
$J_2$ terms decrease while the $J_1$ terms grow when the temperature is lowered.

In order to study the physics at the strong-coupling regime when $J_1$ is of order $1$, we follow Ref. \onlinecite{LSLN} by
employing the Emery-Kivelson unitary transformation $U=\exp{\! \left[i\sqrt{2\pi K}\tilde{\Phi}_s(0)\tilde{S}_z\right]}$,\cite{EK}
where $\tilde{\Phi}_{s/a}=\Phi_{s/a}/\sqrt{K}$ and $\tilde{\Theta}_{s/a}=\sqrt{K}\Theta_{s/a}$. The transformed Hamiltonian
$\tilde{H}=UHU^{\dagger}$ is of the form
\begin{eqnarray}
 \tilde{H} \! \! &=& \! \! H^{(0)}_s+H^{(0)}_a+\lambda\tilde{S}_z\partial_x\tilde{\Theta}_s(0) \nonumber \\
 \! \! & & \! \! +\tilde{S}_x \! \left\{\! \frac{\tilde{J}_{1\perp}}{\pi a_0}\cos{\! \left[\sqrt{2\pi K}\tilde{\Phi}_a(0)\right]}
 \! \! + \! \frac{\tilde{J}_{2\perp}}{\pi a_0}\cos{\! \left[\sqrt{\frac{2\pi}{K}}\tilde{\Theta}_a(0)\right]} \! \right\}
 \nonumber \\
 \! \! & & \! \! +\frac{J_{1E}\tilde{S}_z}{\pi a_0}\sin{\! \left[\sqrt{2\pi K}\tilde{\Phi}_s(0)\right]} \!
 \cos{\! \left[\sqrt{2\pi K}\tilde{\Phi}_a(0)\right]} \! +\delta H , \label{ekh1}
\end{eqnarray}
where $H^{(0)}_{s/a}=\frac{v}{2} \! \int \! \! dx[(\partial_x\tilde{\Theta}_{s/a})^2+(\partial_x\tilde{\Phi}_{s/a})^2]$,
$\lambda=\frac{\tilde{J}_{1z}}{\sqrt{2\pi K}}-\sqrt{2\pi K}v$, $\tilde{J}_{l\perp}=(\tilde{J}_{lx}+\tilde{J}_{ly})/2$,
$\delta\tilde{J}_{l\perp}=\tilde{J}_{lx}-\tilde{J}_{ly}=(J_{l\perp}-J_{lz})\sin^2{\theta}$ ($l=1,2$), and
\begin{eqnarray*}
 \delta H \! &=& \! g_1 \! \left\{\tilde{S}_x\cos{\! \left[\sqrt{8\pi K}\tilde{\Phi}_s(0)\right]}-\tilde{S}_y
 \sin{\! \left[\sqrt{8\pi K}\tilde{\Phi}_s(0)\right]}\right\} \\
 \! & & \! \times\cos{\! \left[\sqrt{2\pi K}\tilde{\Phi}_a(0)\right]} \!
 +g_2\cos{\! \left[\sqrt{\frac{2\pi}{K}}\tilde{\Theta}_a(0)\right]} \\
 \! & & \! \times \! \left\{\tilde{S}_x\cos{\! \left[\sqrt{8\pi K}\tilde{\Phi}_s(0)\right]}-\tilde{S}_y
 \sin{\! \left[\sqrt{8\pi K}\tilde{\Phi}_s(0)\right]}\right\} \\
 \! & & \! +g_3\tilde{S}_z\sin{\! \left[\sqrt{2\pi K}\tilde{\Phi}_s(0)\right]} \!
 \cos{\! \left[\sqrt{\frac{2\pi}{K}}\tilde{\Theta}_a(0)\right]} \\
 \! & & \! +g_4\tilde{S}_z\sin{\! \left[\sqrt{2\pi K}\tilde{\Phi}_a(0)\right]} \!
 \sin{\! \left[\sqrt{\frac{2\pi}{K}}\tilde{\Theta}_a(0)\right]} \\
 \! & & \! +g_5 \! \left\{\tilde{S}_y\cos{\! \left[\sqrt{2\pi K}\tilde{\Phi}_s(0)\right]} \! +\tilde{S}_x
 \sin{\! \left[\sqrt{2\pi K}\tilde{\Phi}_s(0)\right]}\right\} \\
 \! & & \! \times\partial_x\tilde{\Theta}_s(0)+g_6\sin{\! \left[\sqrt{2\pi K}\tilde{\Phi}_a(0)\right]} \!
 \sin{\! \left[\sqrt{\frac{2\pi}{K}}\tilde{\Theta}_a(0)\right]} \\
 \! & & \! \times \! \left\{\tilde{S}_y\cos{\! \left[\sqrt{2\pi K}\tilde{\Phi}_s(0)\right]}+\tilde{S}_x
 \sin{\! \left[\sqrt{2\pi K}\tilde{\Phi}_s(0)\right]}\right\} .
\end{eqnarray*}
In the above, $g_1=\frac{\delta\tilde{J}_{1\perp}}{2\pi a_0}$, $g_2=\frac{\delta\tilde{J}_{2\perp}}{2\pi a_0}$,
$g_3=\frac{J_{2E}}{\pi a_0}$, $g_4=-\frac{\tilde{J}_{2z}}{\pi a_0}$, $g_5=\frac{J_{1E}}{\sqrt{2\pi K}}$,
and $g_6=-\frac{J_{2E}}{\pi a_0}$.

At the vicinity of the point $\lambda=\tilde{J}_{lx}=\tilde{J}_{ly}=\tilde{J}_{lz}=J_{lE}=0$ ($l=1,2$), we may calculate
the scaling dimensions of the various $J$ terms, and the results are shown in the first column in Table \ref{T1}.
When $1/4<K<1$, among the possible relevant $\tilde{J}_{1\perp}$, $\tilde{J}_{2\perp}$, $J_{1E}$, and $g_1$ terms,
the $\tilde{J}_{1\perp}$ term is the most relevant one. At low temperatures, we expect that the effective Hamiltonian
of $\tilde{H}$ is described by the following fixed-point Hamiltonian
\begin{equation}
 H_*=H_s^{(0)}+H_a^{(0)}+\frac{\tilde{J}_{1\perp}}{\pi a_0}\tilde{S}_x\cos{\! \left[\sqrt{2\pi K}\tilde{\Phi}_a(0)\right]}
 . \label{ekh2}
\end{equation}

\begin{table}
\begin{center}
\begin{tabular}{|c|c|c|c|} \hline
 Op.$\backslash$FPs & $\lambda=0=\tilde{J}_{1x(y)}$ & $2$CK & $2$CK$^{\prime}$ \\ \hline
 $\tilde{J}_{1\perp}$ & $K/2$ & N/A & N/A \\ \hline
 $\tilde{J}_{2\perp}$ & $1/(2K)$ & $1/K$ & $1/K$ \\ \hline
 $J_{1E}$ & $K$ & $K/2+1/(2K)$ & $5/(8K)$ \\ \hline
 $\lambda$ & $1$ & $1+1/(2K)$ & $1+5/(8K)$ \\ \hline
 $g_1$ & $5K/2$ & $2K$ & $5/(8K)$ \\ \hline
 $g_2$ & $2K+1/(2K)$ & $2K+1/K$ & $1/K$ \\ \hline
 $g_3$ & $K/2+1/(2K)$ & $K/2+3/(2K)$ & $13/(8K)$ \\ \hline
 $g_4$ & $K/2+1/(2K)$ & $3/(2K)$ & $13/(8K)$ \\ \hline
 $g_5$ & $K/2+1$ & $K/2+1$ & $1$ \\ \hline
 $g_6$ & $K+1/(2K)$ & $K/2+1/K$ & $1/K$ \\ \hline
 $\bar{\lambda}_1$ & N/A & $2/K$ & $2/K$ \\ \hline
 $\bar{\lambda}_2$ & N/A & N/A & $1/(2K)$ \\ \hline
\end{tabular}
\caption{Scaling dimensions of the operators (Op.) at different fixed points (FPs), except those at the Gaussian fixed
         point, which are indicated in the text. $2$CK refers to the fixed point with $\lambda=0$,
         $|\tilde{J}_{1\perp}|\rightarrow +\infty$, and $2$CK$^{\prime}$ refers to the fixed point with $\lambda=0$,
         $|\tilde{J}_{1\perp}|,|\delta\tilde{J}_{1\perp}|\rightarrow +\infty$. $\tilde{J}_{2i}=0$ with $i=x,y,z$ for
         all fixed points. Only the coefficients of corresponding operators are shown.} \label{T1}
\end{center}
\end{table}

We notice that $H_*$ is exactly the 2CK fixed-point Hamiltonian which has been thoroughly studied in Ref. \onlinecite{LSLN}.
Since $[\tilde{S}_x, H_*]=0$, we may set $\tilde{S}_x$ to be its eigenvalues $\pm 1/2$, yielding
$H_{\pm}=H_*(\tilde{S}_x=\pm1/2)$. In $H_{\pm}$, the symmetric sector is described by a spinless LL, while the antisymmetric
sector is described a spinless LL with an impurity backscattering term at $x=0$ and the effective LL parameter $K/2$.
The cosine term is relevant for $K<1$,\cite{KF} which cuts the antisymmetric sector into two separated pieces at $x=0$ at
zero temperature. At finite temperature, tunneling between the two half-wires is allowed, leading to the perturbations
$\bar{\lambda}_1\hat{O}_1$, where $\hat{O}_1=\Psi_A^{\dagger}(0)\Psi_B(0)+{\mathrm H.c.}$ and $\Psi_{A(B)}$ referred to
fermions in the two separated pieces of the wire. The stability of this fixed point in the present case can be examined by
calculating the scaling dimensions of the various perturbations around it in a way similar to that given in Refs.
\onlinecite{LSLN} and \onlinecite{KF}, and the results are shown in the second column in Table \ref{T1}. We see that for
$1/2<K<1$ all perturbations are irrelevant around $H_*$, and this guarantees the stability of the 2CK fixed point in this
region even in the presence of the Rashba interaction.

On the other hand, for $1/4<K<1/2$, the $g_1$ term becomes relevant, and it renders $2$CK fixed point unstable. Since the
$\tilde{J}_{1\perp}$ term also flows to strong coupling in this region, we expect that at low temperatures the system is
described by the new fixed-point Hamiltonian
\begin{eqnarray}
 \tilde{H}_* &=& H_s^{(0)} \! +\frac{\delta\tilde{J}_{1\perp}\xi_a}{2\pi a_0}\tilde{S}_x
 \cos{\! \left[\sqrt{8\pi K}\tilde{\Phi}_s(0)\right]} \nonumber \\
 & & +H_a^{(0)} \! +\frac{\tilde{J}_{1\perp}}{\pi a_0}\tilde{S}_x
 \cos{\! \left[\sqrt{2\pi K}\tilde{\Phi}_a(0)\right]} , \label{ekh3}
\end{eqnarray}
where $\xi_a=\! \left\langle\cos{\! \left[\sqrt{2\pi K}\tilde{\Phi}_a(0)\right]}\right\rangle$. We refer to this new
fixed point as the $2$CK$^{\prime}$ one in the following. Again, $[\tilde{S}_x,\tilde{H}_*]=0$, and we may set
$\tilde{S}_x=\pm 1/2$ so that $\tilde{H}_{\pm}=\tilde{H}_*(\tilde{S}_x=\pm1/2)$. Both the symmetric and the antisymmetric
sectors in $\tilde{H}_{\pm}$ amount to a spinless LL with an impurity backscattering term at $x=0$ and the corresponding
effective LL parameters $K_s=2K$ and $K_a=K/2$, respectively. For $1/4<K<1/2$, the $\tilde{J}_{\perp}$ and the
$\delta\tilde{J}_{1\perp}$ terms both flow to the strong-coupling regime and eventually cut both sectors into two
separated pieces at $x=0$. At finite temperature, tunneling between the two half wires is allowed, leading to the
perturbations $\bar{\lambda}_1\hat{O}_1$ and $\bar{\lambda}_2\hat{O}_2$, where
$\hat{O}_2=\tilde{\Psi}_A^{\dagger}(0)\tilde{\Psi}_B(0)+{\mathrm H.c.}$ and $\tilde{\Psi}_{A(B)}$ referred to fermions in
the two separated pieces of the wire for the symmetric sector. The scaling dimensions of the various perturbations around
the $2$CK$^{\prime}$ fixed point is given in the third column in Table \ref{T1}. We see that all the perturbations are
irrelevant for $1/4<K<1/2$, except the $g_5$ term, which is marginal. This confirms that $\tilde{H}_*$ is indeed the
low-energy effective Hamiltonian in this region.

\section{Physical properties}
One of the most remarkable features of the multichannel Kondo effect is the existence of
the fractionally degenerate ground state, which reveals itself in the impurity entropy at zero temperature. To compute
it, we notice that $\tilde{S}_x$ commutes with the fixed-point Hamiltonian, and thus we may write the partition
function $Z$ as $Z=Z_++Z_-$ for $1/2<K<1$ and $Z=\tilde{Z}_++\tilde{Z}_-$ for $1/4<K<1/2$, where
$Z_{\pm}=\mbox{tr}\{e^{-\beta H_{\pm}}\}$ and $\tilde{Z}_{\pm}=\mbox{tr}\{e^{-\beta\tilde{H}_{\pm}}\}$, respectively.
Following the same reasoning used in Ref. \onlinecite{LSLN}, it can be shown that $Z_+=Z_-$ and $\tilde{Z}_+=\tilde{Z}_-$.
The impurity entropy of $H_+$ ($\tilde{H}_+$) at $T=0$ has been calculated in Ref. \onlinecite{FLS}. It is
$\ln{\sqrt{K/2}}$ for the antisymmetric sector in both $H_+$ and $\tilde{H}_+$ and $\ln{\sqrt{2K}}$ for the symmetric
sector in $\tilde{H}_+$. Together with the contribution from $H_-$ (or $\tilde{H}_-$), we find that $S_{imp}=\ln{(2K)}/2$ for
$1/2<K<1$ and $S_{imp}=\ln{(2K)}$ for $1/4<K<1/2$.

The impurity correction to the free energy is given by
$\delta F_{imp}\equiv F-F_0=-\frac{\lambda^2}{2} \! \int^{\beta}_0 \! \! d\tau C_2(\tau)+O(\lambda^3)$, where $F_0$ is
the bulk free energy at the fixed point,
$C_2(\tau)=\langle {\mathcal T}\{\delta H(\tau)\delta H(0)\}\rangle_0\propto \! \left[\frac{\pi/\beta}{\sin{(\pi\tau/\beta)}}\right]^{2\Delta}$
with $\langle\cdots\rangle_0$ being the expectation value at the fixed point, $\delta H$ denotes the leading irrelevant
operator (LIO), $\lambda$ is the corresponding coupling constant, and $\Delta$ is the scaling dimension of $\delta H$.\cite{FG2}
Near the $2CK$ fixed point, the LIO is the $\frac{J_{1E}\xi_a}{\pi a_0}\tilde{S}_z\sin{\! \left[\sqrt{2\pi K}\tilde{\Phi}_s(0)\right]}$
term with scaling dimension $K/2+1/(2K)$ for $1/\sqrt{3}<K<1$ and the
$\frac{\delta\tilde{J}_{1\perp}\xi_a}{2\pi a_0}\tilde{S}_x\cos{\! \left[\sqrt{8\pi K}\tilde{\Phi}_s(0)\right]}$ term with
scaling dimension $2K$ for $1/2<K<1/\sqrt{3}$. (We notice that in the absence of the Rashba coupling, the LIO is the
$\frac{\tilde{J}_{2\perp}}{\pi a_0}\tilde{S}_x\cos{\! \left[\sqrt{\frac{2\pi}{K}}\tilde{\Theta}_a(0)\right]}$ term with
scaling dimension $1/K$.) On the other hand, near the $2$CK$^\prime$ fixed point, the LIO is the $\bar{\lambda}_2$ term with
scaling dimension $1/(2K)$. From the above results, we may obtain the temperature dependence of the impurity specific heat
$C_{imp}=-T\frac{\partial^2}{\partial T^2}\delta F_{imp}$ at low temperatures, yielding $C_{imp}\propto T^{K+1/K-2}$ for
$1/\sqrt{3}<K<1$, $T^{4K-2} $ for $1/2<K<1/\sqrt{3}$, $T^{1/K-2}$ for $1/3<K<1/2$, and $T$ for $1/4<K<1/3$.

If we apply a small bias across the two edges, a current will flow from one edge to the other. Since only the $J_2$ terms
will contribute to this current, the leading temperature dependence of the conductance $G$ at zero bias reflects the
renormalization-group (RG) flow of the $\tilde{J}_{2\perp}$ term. From the above discussions, we see that neither the
qualitative RG flow of the $\tilde{J}_{2\perp}$ term or its scaling dimensions near the various fixed points are affected
by the the presence of the Rashba coupling. As a result, the temperature dependence of $G$ is identical to that without
the Rashba coupling. [See Fig. $1$(b) in Ref. \onlinecite{LSLN}.]

One way to distinguish the $2$CK and $2$CK$^{\prime}$ fixed point is to investigate the dynamical structure factor of
the impurity spin $\chi_i(\omega, T)$, which can be obtained from
${\mathcal S}_i(i\omega_n)=\! \int^{\beta}_0 \! \! d\tau e^{i\omega_n\tau}S^{(2)}_i(\tau)$ by analytic continuation
$i\omega_n\rightarrow\omega+i0^+$, where $i=y,z$ and $S_i^{(2)}(\tau)=-\langle{\mathcal T}_{\tau}\{S_i(\tau)S_i(0)\}\rangle$.
After performing the unitary transformation $U$, we have
\begin{eqnarray*}
 S_y \! \! \! &=& \! \! \cos{\theta} \! \left\{\sin{\! \left[\sqrt{2\pi K}\tilde{\Phi}_s(0)\right]} \! \tilde{S}_x \!
 +\sin{\! \left[\sqrt{2\pi K}\tilde{\Phi}_s(0)\right]} \! \tilde{S}_y \! \right\} \\
 \! \! & & \! \! +\sin{\theta}\tilde{S}_z \ , \\
 S_z \! \! \! &=& \! \! -\sin{\theta}\! \left\{\sin{\! \left[\sqrt{2\pi K}\tilde{\Phi}_s(0)\right]} \! \tilde{S}_x \!
 +\sin{\! \left[\sqrt{2\pi K}\tilde{\Phi}_s(0)\right]} \! \tilde{S}_y \! \right\} \\
 \! \! & & \! \! +\cos{\theta}\tilde{S}_z \ .
\end{eqnarray*}
Near the fixed point, we expect that $S_i^{(2)}(\tau)\propto |\tau|^{-x_i}$ at $T=0$ as $|\tau|\rightarrow +\infty$,
and $x_i$ is determined by the term in the above equations with the smallest scaling dimension.\cite{Tsv} Near both the
$2$CK and the $2$CK$^{\prime}$ fixed point, it is the $\tilde{S}_x$ term which determines $x_i$, yielding $x_i=K$ for
$1/2<K<1$ and $2K$ for $1/4<K<1/2$. Hence, we find that at low frequencies $\mbox{Re}\chi_i(\omega,0)\approx C_i|\omega|^{K-1}$
for $1/2<K<1$ and $C_i|\omega|^{2K-1}$ for $1/4<K<1/2$, and $\mbox{Re}\chi_i(0,T)\approx\tilde{C}_iT^{K-1}$ for $1/2<K<1$
and $\tilde{C}_iT^{2K-1}$ for $1/4<K<1/2$ at low temperatures, where $C_y,\tilde{C}_y\propto\cos^2{\theta}$ and
$C_z,\tilde{C}_z\propto\sin^2{\theta}$.

\section{Quantum critical regime}
From the above analysis, the difference between the $2$CK and the $2$CK$^{\prime}$ phases
lies at the Hamiltonian for the symmetric sector. Hence, at the critical point separating these two phases, the system can be
described by the Hamiltonian for the
symmetric sector with the LL parameter $K=1/2$. It turns out that this Hamiltonian can be refermionized as
\begin{equation}
 H_s \! =v \! \int \! \! dx \! \left(\psi_l^{\dagger}i\partial_x\psi_l-\psi_r^{\dagger}i\partial_x\psi_r\right) \! +m \!
 \left[i\psi_l^{\dagger}\psi_r(0)+{\mathrm H.c.}\right] . \label{ekh41}
\end{equation}
where $m=\pm\delta\tilde{J}_{1\perp}\xi_a/2$ and $\psi_{l/r}\propto e^{\mp i\sqrt{\pi}(\tilde{\Phi}_s\pm\tilde{\Theta}_s)}$.
In Eq. (\ref{ekh41}), we have set $\tilde{S}_x=\pm 1/2$ because $[\tilde{S}_x,H_s]=0$. $H_s$ is nothing but the Hamiltonian
of a one-dimensional Fermi liquid backscattered by a $\delta$-function like potential at $x=0$. Therefore, we expect that
its thermodynamical properties should resemble those of the Fermi liquid. In fact, straightforward calculations show that
$S^{(s)}_{imp}=0~(T=0)$ and $C^{(s)}_{imp}\propto T$. Since the combined contribution to $S_{imp}$ from the antisymmetric
sector and $\tilde{S}_x$ vanishes, we conclude that $S_{imp}=0$ at the QCP. Moreover, the contribution to $C_{imp}$ arising
from the antisymmetric sector is given by the LIO, which will give a higher power in $T$. Consequently, the leading temperature
dependence of $C_{imp}$ is dominated by the symmetric sector, i.e. $C_{imp}\propto T$. Logarithmic corrections to this result
are possible and can be obtained by the one-loop RG equation near the critical point $K=1/2$. However, this is beyond the
scope of the present work.

\section{Concluding remarks}
To summarize, we have shown that although the low-temperature physics of a QD coupled to two
helical edge states is still described by the $2$CK fixed point
in the presence of a Rashba coupling for $1/2<K<1$, the
leading temperature dependence of thermodynamical quantities is drastically changed due to the new LIO's produced by the
Rashba interaction. With stronger Coulomb repulsions between electrons on the edges, the system will be driven to the $2$CK$^{\prime}$ phase.
This phase is characterized by a new set of fixed point (line) Hamiltonians,
and it can be distinguished from the $2$CK phase by examining the impurity
spin susceptibilities. At the boundary between the two phases, the system exhibits scaling behaviors which are
distinct from those in the 2CK and 2CK$^\prime$ phases, as we have shown. Since the LL parameter $K$
depends sensitively on the Rashba interaction strength, the new phase described by the $2$CK$^{\prime}$ fixed point and the
quantum phase transition into this phase can be electrically controlled. Therefore, our results not only reveal a new $2$CK
fixed point that has not been analyzed before, but also serve as a useful guide for future experimental investigations on this
system.

\acknowledgments

The work of Y.-W. Lee is supported by the National Science Council of Taiwan under Grant No. NSC 99-2112-M-029-006-MY3.



\begin{thebibliography}{99}

 \bibitem{Konig} M. K\"{o}nig et al., Science {\bf 318}, 766 (2007). 
 \bibitem{BHZ} B.A. Bernevig, T.L. Hughes, and S.C. Zhang, Science {\bf 314}, 1757 (2006). 
 \bibitem{Qi_RMP} For a review, see X.-L. Qi and S.-C. Zhang, Rev. Mod. Phys. {\bf 83}, 1057 (2011).
 \bibitem{GuWen} Z.C. Gu, X.G. Wen, Phys. Rev. B {\bf 80}, 155131 (2009).
 \bibitem{KM} C.L. Kane and E.J. Mele, Phys. Rev. Lett. {\bf 95}, 226801 (2005).
 \bibitem{BZ} B.A. Bernevig and S.C. Zhang, Phys. Rev. Lett. {\bf 96} 106802 (2006).
 \bibitem{KM2} C.L. Kane and E.J. Mele, Phys. Rev. Lett. {\bf 95}, 146802 (2005).
 \bibitem{WBZ} C. Wu, B.A. Bernevig, and S.C. Zhang, Phys. Rev. Lett. {\bf 96}, 106401 (2006); C. Xu and J.E. Moore,
         Phys. Rev. B {\bf 73}, 045322 (2006).

 \bibitem{SI} A. Schiller and K. Ingersent, Phys. Rev. B {\bf 51}, 4676 (1995).
 \bibitem{MLOQWZ} J. Maciejko, C. liu, Y. Oreg, X.L. Qi, C. Wu, and S.C. Zhang, Phys. Rev. Lett. {\bf 102}, 256803
         (2009).
 \bibitem{TFM} Y. Tanaka, A. Furusaki, and K.A. Matveev, Phys. Rev. Lett. {\bf 106}, 236402 (2011).

 \bibitem{ESSJ} E. Eriksson, A. Str\"{o}m, G. Sharma, and H. Johannesson, Phys. Rev. B {\bf 86}, 161103(R) (2012).

 \bibitem{LSLN} K.T. Law, C.Y. Seng, P.A. Lee and T.K. Ng, Phys. Rev B {\bf 81}, 041305(R) (2010).

 \bibitem{GR} L.I. Glazman and M.E. Raikh, JETP Lett. {\bf 47}, 452 (1988).
 \bibitem{NL} T.K. Ng and P.A. Lee, Phys. Rev. Lett. {\bf 61}, 1768 (1988).


 \bibitem{FG} M. Fabrizio and A. O. Gogolin, Phys. Rev. B {\bf 51}, 17827 (1995).

 \bibitem{Wi} R. Winkler, {\it Spin-Orbit Interaction Effects in Two-Dimensional Electron and Hole Systems}
 (Springer, Berlin, 2003).
 \bibitem{Bu} H. Buhmann, J. Appl. Phys. {\bf 109}, 102409 (2011).

 \bibitem{Hew} C. Hewson, {\it The Kondo Problem to Heavy Fermions} (Cambridge University Press, Cambridge, 1993).

 \bibitem{ZPP} R. \v Zitko, R. Peters, and Th. Pruschke, Phys. Rev. B {\bf 78}, 224404 (2008).
 \bibitem{VO} J.I. V\"{a}yrynrn and T. Ojanen, Phys. Rev. Lett. {\bf 106}, 076803 (2011).

 \bibitem{GNT} A.O. Gogolin, A.A. Nersesyan, and A.M. Tsvelik, {\it Bosonization and Strongly Correlated Systems}
         (Cambridge University Press, Cambridge, U.K., 1999).


 \bibitem{EK} V.J. Emery and S. Kivelson, Phys. Rev. B {\bf 46}, 10812 (1992).

 \bibitem{KF} C.L. Kane and M.P.A. Fisher, Phys. Rev. B {\bf 46}, 15233 (1992).
 \bibitem{FLS} P. Fendley, F. Lesage, and H. Saleur, J. Stat. Phys. {\bf 79}, 799 (1995).

 \bibitem{FG2} M. Fabrizio and A. O. Gogolin, Phys. Rev. B {\bf 50}, 17732 (1994).
 \bibitem{Tsv} A.M. Tsvelick, J. Phys.: Condens. Matter {\bf 2}, 2833 (1990).


\end{thebibliography}
\end{document}